\documentclass[preprint,authoryear,showpacs,showkeys]{revtex4}
\usepackage{footmisc}
\DefineFNsymbols{mySymbols}{{\ensuremath*}{\ensuremath{**}}\S\P
   *{**}{\ensuremath{\dagger\dagger}}{\ensuremath{\ddagger\ddagger}}}
\setfnsymbol{mySymbols}
\usepackage{amssymb,amsmath}
\usepackage{graphicx}
\usepackage{tensor}
\usepackage[varg]{txfonts}
\usepackage{amsbsy}

\begin{document}

\title{On a correlation among azimuthal velocities and the flyby anomaly sign}
\author{L. Acedo}
\email{luiacrod@imm.upv.es}
\author{Ll. Bel}
\thanks{Retired}
\affiliation{Instituto Universitario de Matem\'atica Multidisciplinar, Building 8G, 2$^\circ$ floor,
Universitat Polit\`ecnica de Val\`encia, Camino de Vera s/n, E-46022, Valencia, Spain}

\date{\today}

\begin{abstract}

Data of six flybys, those of Galileo I, Galileo II, NEAR, Cassini, Rosetta and Messenger
were reported by Anderson et al \citep{Anderson}. Four of them: Galileo I, NEAR, Rosetta  and Messenger
gain Newtonian energy during the flyby transfer, while Galileo II and Cassini lose energy.
This is, in both cases, a surprising anomaly since Newtonian forces derive from a potential
and they are, therefore, conservative. We show here that the gravitational field of a rotating planet as derived from a new model introduces a non conservative force that gives a partial, but in our opinion satisfactory, explanation of these anomalies and suggests a correlation between the sign of the anomaly
and the sign of the azimuthal velocity at perigee.

\end{abstract}


\maketitle

\section{Introduction}
\label{intro}

Extensions of Whitehead's and Whitehead-Synge's models \citep{Whitehead,Coleman} have been considered by one of us in two occasions \citep{Bel1} and \citep{Bel2}. The first
extended model was still, in many respects, equivalent to the very first approximation of General relativity. The second extension used here
includes, from the beginning, the retardation effects taking into account that gravitational fields propagate at the speed of the universal constant $c$,
and is, in many respects, equivalent to the first approximation of General relativity when retardation is taking into account. This approach had already been considered by Hafele \citep{Hafele} but somewhere it departed from ours.

Our model lead to a generalization of Newton's theory where {\it post-Newtonian pre-relativistic corrections of} order $1/c$ are relevant, as well as the relativistic corrections of order $1/c^2$. This model has been used before by one of us \citep{Acedo} using perturbation theory and a particular set of azimuthal velocities at the perigee to calculate a first estimation of the anomalies.

More generally we describe here the gravitational field of the Earth as a spherically symmetric Newtonian field, slightly modified by a quadrupole
contribution, and a novel contribution due to its rotation. The corresponding differential equations describing the motion of any aircraft by-flying the Earth being \citep{Bel2}:

\begin{eqnarray}
\frac{d^2x}{dt^2}&=&-\frac{GM}{r^3}x-\frac32 \frac{J_2 x (x^2+y^2-4 z^2)}{r^7}+\xi\frac{GM R^2\Omega}{5 c r^4}y \nonumber \\
\noalign{\smallskip}
\frac{d^2y}{dt^2}&=&-\frac{GM}{r^3}y-\frac32 \frac{J_2 y (x^2+y^2-4 z^2)}{r^7}-\xi\frac{GM R^2\Omega}{5 c r^4}x  \nonumber \\
\noalign{\smallskip}
\label{Diffs}
\frac{d^2z}{dt^2}&=&-\frac{GM}{r^3}z-\frac32 \frac{J_2 z (x^2+y^2-4 z^2)}{r^7}
\end{eqnarray}
where $ r=\sqrt{x^2+y^2+z^2}$, and $G M$ is the gravitational constant times the mass of the Earth, $\Omega$ is the Earth's angular velocity of rotation around its axis, $J_2$ is the lowest order zonal harmonic measuring the ellipticity of the Earth and $c$ is the speed of light whose values are:

\begin{eqnarray}
\label{GM}
 GM=398600.440\, \mbox{km$^3$\,s$^{-2}$} &\, , & \Omega=+7.292115\, 10^{-5}\,\mbox{s$^{-1}$} \, , \\
\noalign{\smallskip}
\label{J2}
J_2=1.7555\,10^{10}\, \mbox{km$^5$\, s$^{-2}$} & \, , & c=299792.4580\, \mbox{km\,s$^{-1}$}\, .
\end{eqnarray}

Here $\xi$ includes, in general, a correction due to the inhomogeneity of the spherical mass density \citep{Hafele}. $\xi$ is, approximately, $1$ and this is the value we will consider in this paper. Note that no phenomenological parameter has been included, and this is an essential difference with all other models we know of, including Hafele's \citep{Hafele}. In this paper we show that the azimuthal velocity at perigee is correlated with the sign of the anomalous energy change as discussed by Anderson et al. \citep{Anderson}, i. e., for spacecrafts flybying the Earth opposite to its rotation an anomalous energy increase is expected but if the flyby is performed in the same direction as Earth's rotation we predict an energy decrease. The predictions of the model agree with this correlation as well as the observed anomalies with the exception of the Galileo II flyby which, on the other hand, provided data of low reliability because the
effect of atmospheric friction was important in this particular flyby.

The paper is organized as follows: In Section \ref{secvh} we discuss a method to obtain the azimuthal velocity at perigee from a combination of the asymptotic velocities and the perigee's radiovector. Integration results for the anomalous energy changes are given in Section \ref{secint}. A discussion about the results and some conclusions are given in Section \ref{secdiscuss}.
Further details on the proposed Whitehead's model is provided in Appendix \ref{apenmodel}.

\section{The azimuthal velocity at the flyby's perigee}
\label{secvh}

As it is the case in the perturbation approach to integrate the system of equations above the knowledge of initial conditions at perigee is necessary. The distance $h$ from the center of the Earth to the perigee and the speed $v$ at the perigee for each of the six flybys are listed in \citep{Anderson,Acedo}, as well as the longitude $\lambda$ and latitude $\phi$. 
But this information is not sufficient and the values of the azimuthal component $va$ of the velocity along the circle of constant latitude passing through the perigee are also necessary.
They are also listed in Ref. \citep{Anderson}, using a celestial equatorial system of reference, the declination $\delta_i$ and right ascension $\alpha_i$ of the incoming asymptotic direction of the flyby  as well as the corresponding asymptotic outgoing quantities  $\delta_o$ and  $\alpha_o$.

Below we propose a simple method to derive a reasonable value $va$. 

It uses the data above as follows: lets define two vectors with origin at the perigee and components:

\begin{eqnarray}
\label{exs}
\vec ve_i:\ ex_i&=&\cos(\delta_i)\cos(\alpha_i), \\
\noalign{\smallskip}
 \ ey_i&=&\cos(\delta_i)\sin(\alpha_i), \\
\noalign{\smallskip}
 \ ez_i&=&\sin(\delta_i) \; ,
\end{eqnarray}
and similar equations for the components of  $\vec ve_o$ in terms of $\delta_o$ and $\alpha_o$.
The plane that these two vectors define is the orbital plane and its inclination on the ecliptic is:

\begin{eqnarray}
I=\arccos\left(\frac{ex_i\, ey_o-ey_i\, ex_o}{|\vec ve_i \times \vec ve_o|}\right)
\end{eqnarray}
The corresponding values of $I$ are listed in \citep{Anderson} as the inclination of the orbital plane on the Earth equator.

We have found convenient to define the two orthogonal unit vectors:

\begin{eqnarray}
\label{u,w}
\vec ve^+=\frac{\vec ve_o+\vec ve_i}{|\vec ve_o+\vec ve_i|}; & \; , \quad & \vec ve^-=\frac{\vec ve_o-\vec ve_i}{|\vec ve_o-\vec ve_i|}\; .
\end{eqnarray}

On the other hand, the components  of the position vector of the perigee are:

\begin{eqnarray}
\label{rs}
\vec h:\ hx&=&h\cos(\delta_p)\cos(\alpha_p), \\
\noalign{\smallskip}
 hy&=&h\cos(\delta_p)\sin(\alpha_p), \\
\noalign{\smallskip}
hz&=&h\sin(\delta_p)
\end{eqnarray}

where $\delta_p$, $\alpha_p$ are the declination and the right ascension of the perigee.

With these data we require the unit velocity $\vec vp$ at the perigee to be:

\begin{equation}
\label{vec v}
\vec vp=vp(\cos(\zeta)\vec vq^++\sin(\zeta)\vec vq^-)
\end{equation}
where $vp$ is the speed at perigee, listed in \citep{Anderson}, and $\zeta$ is one of the two solutions of the equation:

\begin{equation}
\label{condition}
\vec vp\centerdot\vec h=0
\end{equation}
Obviously if $\zeta$ is a solution then $\zeta+\pi$ is the second solution.
Now, the azimuthal component is given by:
\begin{equation}
\label{parallel}
va=-vx\sin(\alpha_p)+vy\cos(\alpha_p) \; .
\end{equation}
But to calculate it with the right sign we must discriminate among the two solutions of Eq. (\ref{condition}). To do so we define the
positive direction of the azimuthal velocity as that of the rotation of the Earth as seen from the polestar, i. e., the counterclokwise 
direction. If we notice that the projection of the spacecraft's radiovector goes from the incoming direction to the outgoing direction passing through the perigee we have that the sign of the acute angle formed by the incoming and outgoing directions is that of $va$.

The results for six flybys discussed in \citep{Anderson} are listed in Table \ref{tabI}.
\begin{table}
 \centering
\caption{Azimuthal velocities (km/sec) at perigee for six flybys.}
\label{tabI}
\begin{tabular}{cccc}\hline
 & \bf{Galileo I} & \bf{Galileo II} & \bf{NEAR}\\
\hline
va & -11.993 & -12.729 & -4.694 \\
\hline
 & \bf{Cassini} & \bf{Rosetta} & \bf{Messenger} \\
va & 18.740 & -9.170 & -10.387 \\
\hline
\end{tabular}
\end{table}
For the Juno flyby performed on October, $9^{\mbox{th}}$, 2013 we obtain $va=10.600$ km/sec. So, only the Cassini and Juno flybys were performed in the same direction as that of the rotation of the Earth around its axis. In the rest (Galileo I, Galileo II, NEAR, Rosetta and Messenger) the azimuthal velocity at perigee was opposite to that of the Earth's rotation.

\section{Integration results}
\label{secint}
We have integrated numerically the system of differential equations using as initial conditions at the perigee the components $px,py,pz$ of the position vector and the corresponding components of the velocity $vx,vy,vz$, for an interval of time $\Delta t=20\, hours$,  from $t_i=10\,hours$ before the perigee to $t_o=10\,hours$ after the perigee.

More precisely the relevant information that we get from each integration are the dimensionless quantity:

\begin{equation}
\label{Anomaly}
Qf=\sqrt{E(t_o)/E(t_i)}-1;
\end{equation}
where:

\begin{equation}
\label{Qf}
E=\frac12 v^2-\frac{GM}{r}+\frac12 \frac{J_2 (-r^2+3 z^2)}{r^5}
\end{equation}
would be the Energy of the spacecraft in the gravitational field if $\Omega$ were zero, and

\begin{equation}
\label{dv}
\delta v=v(t_o)-v(t_i)
\end{equation}
where $v$ is the speed of the aircraft.

Without the terms proportional to $\Omega$  in Eq. (\ref{Diffs}), $E$ would be a constant of motion and both quantities $Qf$ and $\delta v$
would vanish. On the contrary, if one includes $\Omega$, then $E$ no longer is a constant of motion and $Qf$ and $\delta v$ have values different from zero that we believe correspond to the otherwise known as the anomalies of the Flybys.

In Table \ref{tabII} we list the values of $Qf$ and $\delta v\, \mbox{mm s$^{-1}$}$ taking into account the
ellipticity term ($J_2 \neq 0$) and in the case of a perfectly spherical planet ($J_2=0$). $\Delta v$ is the observed value of $\delta v$ as given in \citep{Anderson}.

\begin{table}
 \centering
\caption{Integration results for the flyby anomaly in terms of $Qf$ and the anomalous variation of asymptotic velocities, $\delta v$ for an ideal spherical Earth and for $J_2$ given in Eq. (\protect\ref{J2}) compared with the results by Anderson et al., $\Delta v$, as given in Ref. \protect\citep{Anderson}.}
\label{tabII}
\begin{tabular}{cccc}\hline
 & \bf{Galileo I} & \bf{Galileo II} & \bf{NEAR}\\
\hline
$10^7$ $Qf$ ($J_2=0$) & 2.741 & 3.229 & 2.180 \\
$10^7$ $Qf$ ($J_2 \neq 0$) & 2.742 & 3.229 & 2.180 \\
$\delta v$ ($J_2=0$) & 2.387 & 2.789 & 1.414 \\
$\delta v$ ($J_2 \neq 0$) & 2.687 & 2.417 & 3.211 \\
$\Delta v$  & 3.92 & -4.56 & 13.46 \\
\hline
 & \bf{Cassini} & \bf{Rosetta} & \bf{Messenger} \\
$10^7$ $Qf$ ($J_2=0$) & -0.834 & 13.507 & 9.188 \\
$10^7$ $Qf$ ($J_2 \neq 0$) & -0.834 & 13.507 & 9.178 \\
$\delta v$ ($J_2=0$) & -1.329 & 4.234 & 3.094 \\
$\delta v$ ($J_2 \neq 0$) & -1.338 & 6.380 & 3.207 \\
$\Delta v$  & -2 & 1.8 & 0.02 \\
\hline
\end{tabular}
\end{table}

Notice that the agreement (in order of magnitude and sign) is good in five cases but the predicted sign is positive for the Galileo II flyby for which a total decrease of $8$ mm/sec was measured after fitting the postencounter Doppler data. Anderson et al. assumed that the atmospheric drag could be estimated in
$-3.4$ mm/sec giving a residue of $\Delta v \simeq -4.6$ mm/sec. However, no detailed analysis of the atmospheric drag has been provided for the geometry of this spacecraft so a drag sufficiently large to
leave a positive unexplained residue in the asymptotic velocity cannot be discarded. If we compare Tables \ref{tabI} and \ref{tabII} it is clear that a negative azimuthal velocity (corresponding to a spacecraft orbiting the Earth opposite to its rotation) is correlated with a positive anomalous energy change (this is the case for the Galileo I, NEAR, Rosetta and Messenger flybys). On the other hand, for Cassini, which flybyed the Earth in the direction of its rotation, we get an anomalous energy decrease.
This correlation could help in finding an explanation to the anomaly in the context of an extended
theory of gravity and it reinforces the intuition of Anderson et al. \citep{Anderson} as they related the anomaly
to an enhanced frame dragging effect generated by Earth's rotation.

\begin{figure}
\includegraphics[width=\linewidth]{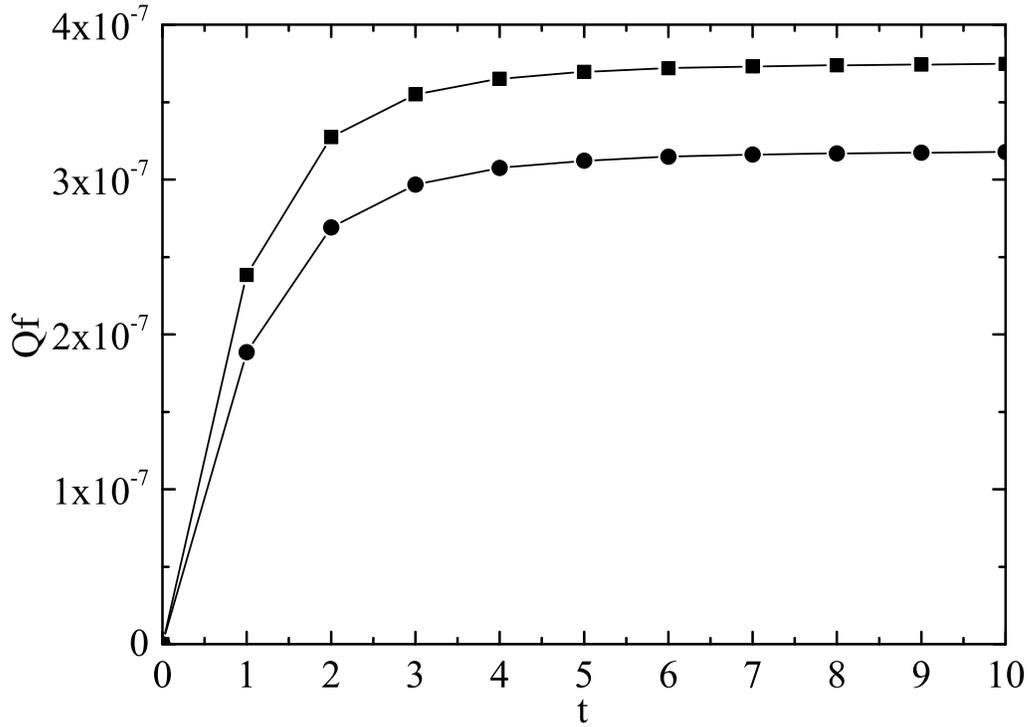}
\caption{Evolution of the fractional anomalous energy changes, according to the Whitehead extended theory, for the Galileo I (circles), Galileo II (squares). Here $t$ is the time in hours from the closest approach.}
\label{fig1}
\end{figure}

\begin{figure}
\includegraphics[width=\linewidth]{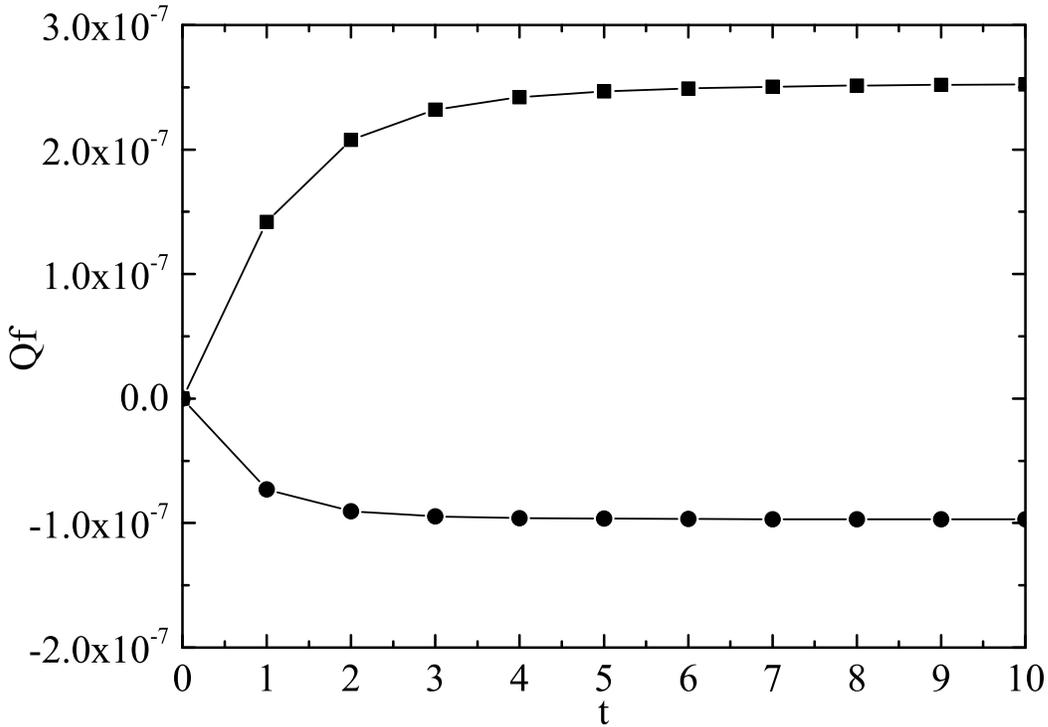}
\caption{The same as Figure \protect\ref{fig1} but for the NEAR (squares) and Cassini (circles) flybys. Notice that in the Cassini flyby the total energy decreases in contrast with the anomalous increase detected in other flybys also predicted by the model.}
\label{fig2}
\end{figure}

\begin{figure}
\includegraphics[width=\linewidth]{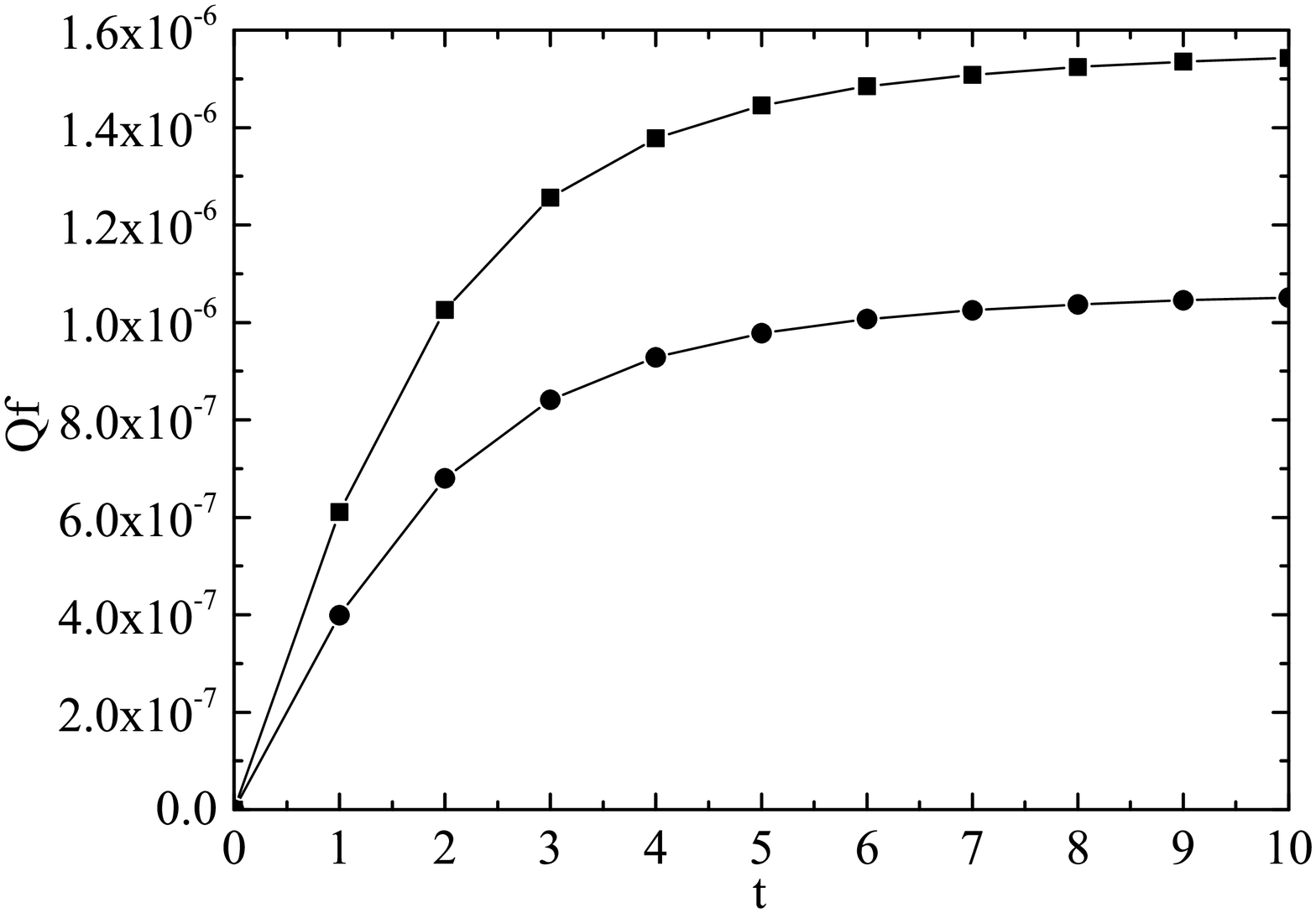}
\caption{The same as Figure \protect\ref{fig1} but for the Messenger (circles) and Rosetta (squares) flybys.}
\label{fig3}
\end{figure}

In Figs. \ref{fig1}, \ref{fig2} and \ref{fig3} we have plotted the evolution of the anomalous energy changes according to the extended Whitehead model given by Eqs. \ref{Diffs} discussed in Appendix \ref{apenmodel}. We notice that most of the anomalous energy change takes place around the perigee. This could be expected from the form of the non-Newtonian interaction because, as shown in Eqs. \ref{Diffs}, it decreases with the third power of the distance to the center of the Earth.

In the case of Juno we obtain $\delta v=-0.52$ mm/sec (for $J_2=0$). This means that the same correlation among the azimuthal velocity at perigee ($va=10.600$ km/sec) and the sign of the anomalous energy variation is verified in the model for this flyby.
To the best of our knowledge this flyby is still not analyzed in the search of a possible flyby anomaly  \citep{IorioJuno,Acedo2} but it will be very interesting to check if the correlation discussed in this paper is supported also by this analysis.

\section{Discussion and Conclusions}
\label{secdiscuss}
\begin{itemize}

\item The first remark to make is that the values of $\delta v$ derived from the model and the values of $\Delta v$ observed are of the same order of magnitude, a few mm/s and this could not be anticipated from a model that does not contains any free parameter.

\item Four flybys: Galileo I, NEAR, Rosetta and Messenger gain energy and two,  Galileo II and Cassini, lose energy. The correlation of this fact with the sign of the azimuthal component of the velocity at perigee, being in the direction opposite to the rotation of the Earth, is correct except for Galileo II. This fact, combined with the low altitude of the flyby suggest, as Anderson et al. \citep{Anderson} already feared, that the data for Galileo II are unreliable.

\item The values of $\delta v$ corresponding to Galileo I and Cassini are good. That of Galileo I improves when the quadrupole of the Earth is included in the model. That of Cassini is unaffected. The values of $\delta v$ corresponding to NEAR, Rosetta and Messenger are not very good and point towards the necessity of considering external perturbations to the Earth inertial system.

\item The graphs of the solutions for GalileoI and Cassini clearly indicate that the transfer of energy leading to the anomaly, while being progressive, increases rapidly in a neighborhood of the perigee. The graphs of the remaining flybys are very similar to that of GalileoI.

\item It follows from the preceding remarks that an accurate description of the gravitational field of the Earth has to include the rotation term in Eqs. \ref{Diffs}, including $\xi$ as a free parameter to be found jointly in a statistic analysis, with none of its multi-poles  considered to be a priori independent of $\xi$.

\item In accordance with this correlation we predict a negative energy change for the, still not analyzed, Juno flyby. This is in contrast with the prediction of Anderson's phenomenological formula \citep{Anderson,IorioJuno} but other models also lead to a negative energy change \citep{Acedo2}. We expect that this work will encourage further studies to confirm or dismiss our proposal for a pattern in the flyby anomaly according to the direction of the spacecraft.

\end{itemize}

\acknowledgments
  We acknowledge JPL Solar System Dynamics Group for providing the data used in this work through the online Horizons ephemerides system.


%
%

\section{Preliminaries to an extended Whitehead model of gravity}
\label{apenmodel}
As it is the case in Whitehead's and Synge's models the gravity model \citep{Bel2}, used in the main body of this paper, considers from
the beginning the retardation effects due to the fact that gravity fields propagate at the speed of the universal constant $c$. But it differs from them in the fact that it remains much closer to some of the innovations brought to us by General relativity, as for example having Einstein's field equations at the center of the formalism. Keep in mind though that our model may be compared only to the linear approximation of General relativity, not to the full theory. On the other hand at the linear approximation it brings the explicit influence of retardation effects of paramount importance in the problem discussed in this paper.

Let us consider a world-line $W:\hat x^\alpha(\tau)$ and let $u^\alpha$ be the unit time-like tangent vector at $\hat x^\alpha$. Let $x^\alpha$ be an event in the future of $\hat x^\alpha(\tau)$ and define $L^\alpha=x^\alpha-\hat x^\alpha$ and $\hat r$ so that:

\begin{equation}
\label{L,u}
L^\alpha L_\alpha=0, \ (L^0>0), \  \hat u^\alpha \hat u_\alpha=-1, \ (u^0>0), \ \hat r=-u_\alpha L^\alpha >0.
\end{equation}
$\hat r$ is the retarded distance from $x^\alpha$ to the world line of the point mass $m$.

Our model starts with a a 4-dimensional quadratic  hyperbolic form:

\begin{equation}
\label{ds2}
ds^2= g_{\alpha\beta}dx^\alpha dx^\beta
\end{equation}
where the potentials can be approximated, as usual in the linear approximation model of General relativity, by a sum:

\begin{equation}
\label{h}
 g_{\alpha\beta}=\eta_{\alpha\beta}+h_{\alpha\beta}(L^\gamma)
\end{equation}
 but differs from it by the fact that the deviations $h_{\alpha\beta}$ are functions of the variables $x^\alpha$ through the vector components of
 $L^\alpha$ that has to remain null when differentiated no matter how many times. This means that, when considering a variation $\delta x^\alpha$,
 $\hat \tau$ has to be displaced accordingly along $W$. This is achieved defining the derivative symbol $\hat \partial$
 by the sequence of conditions:

\begin{equation}
\label{hat delta}
\hat\partial_\alpha L^\beta=\delta^\beta_\alpha+\frac{1}{\hat r}u^\beta L_\alpha,
\  \ \hat \partial_\alpha u^\beta=-\frac{1}{r}\dot u^\beta L_\alpha, \  \ \cdots
\end{equation}
where an overhead dot means a derivative with respect to $\tau$ at the retarded event. We further simplify the model assuming that either all derivatives of $u^\alpha$ are small enough or the two events
$x^\alpha$ and $\hat x^\alpha$ are close enough so that the only derivatives that we have to consider is the first one above.

Whitehead did not define any causality preserving differentiation and neither Whitehead nor Synge cared about stablishing their theories as well defined field theories with field equations. The model used to derive the system of equations is:

\begin{equation}
\label{Ricci}
R_{\alpha\beta}=0, \quad R_{\alpha\beta}=\eta^{\lambda\mu}R_{\alpha\lambda\beta\mu}
\end{equation}
where $R_{\alpha\beta}$ is the linear part of the Ricci tensor. We have also that:

\begin{equation}
\label{Riemann}
R_{\alpha\lambda\beta\mu}=-\frac12
(\hat\partial_{\alpha\lambda}h_{\beta\mu}+\hat\partial_{\beta\mu}h_{\alpha\lambda}
-\hat\partial_{\alpha\mu}h_{\beta\lambda}-\hat\partial_{\beta\lambda}h_{\alpha\mu})=0
\end{equation}
for the linear part of the Riemann tensor. Here we use the partial derivatives $\hat \partial_\alpha$  instead of $\partial_\alpha$.

Finally, the equations of motion of a test particle in the field of a heavy one are:

\begin{equation}
\label{EofM}
\frac{d^2x^\gamma}{ds^2}=-\Gamma^\gamma_{\alpha\beta}\frac{dx^\alpha}{ds}\frac{dx^\beta}{ds}
\end{equation}
where the same substitution as before of $\partial$ by $\hat\partial$ is made in the expressions of the Christoffel symbols.

\end{document}